\documentstyle{aa}
\input epsf.sty

\def\plotone#1{\centering \leavevmode \epsfxsize=\columnwidth \epsfbox{#1}}

\def\apj{ApJ}
\def\aj{AJ}
\def\aap{A\&A}
\def\aj{{\it AJ}}
\def\mnras{MNRAS}
\def\nature{Nature}

\begin{document}
\thesaurus{11   (11.03.4 A426; 11.03.3; 11.09.1 NGC 1275; 11.09.4;
              13.25.2)}
\title{Asymmetric, arc minute scale structures around NGC 1275}

\author{E.Churazov \inst{1,2}
\and W.Forman \inst{3}
\and C.Jones \inst{3}
\and H.B\"{o}hringer \inst{4}}

\institute{MPI f\"{u}r Astrophysik, Karl-Schwarzschild-Str.1, 85740
Garching, Germany
\and Space Research Institute, Profsouznaya 84/32, Moscow 117810, 
Russia
\and Harvard-Smithsonian Center for Astrophysics, 60 Garden St.,
Cambridge, MA 02138
\and MPI f\"{u}r Extraterrestrische Physik, P.O.Box 1603, 85740
Garching, Germany
}

\maketitle

\sloppypar

\begin{abstract}
ROSAT HRI observations show complicated substructure in the X--ray
surface brightness within $\sim$5 arcminutes around NGC 1275 -- the
dominant galaxy of the Perseus cluster. The typical amplitude of the
variations is of the order of 30\% of the azimuthally averaged surface
brightness at a given distance from NGC 1275. We argue that this
substructure could be related to the activity of NGC 1275 in the
past. Bubbles of relativistic plasma, inflated by jets, be forced to
rise by buoyancy forces, mix with the ambient intracluster medium
(ICM), and then spread. Overall evolution of the bubble may resemble
the evolution of a hot bubble during a powerful atmospheric
explosion. From a comparison of the time scale of the bubble inflation
to the rise time of the bubbles and from the observed size of the
radio lobes which displace the thermal gas, the energy release in
the relativistic plasma by the active nucleus of NGC 1275 can be
inferred. Approximate modeling implies a nuclear power output of the
order of $10^{45}$ erg s$^{-1}$ averaged over the last $\sim 3~10^7$
years. This is comparable with the energy radiated in X-rays during
the same epoch.  Detailed measurements of the morphology of the X--ray
structure, the temperature and abundance distributions with Chandra
and XMM may test this hypothesis.
\end{abstract}


\section{INTRODUCTION}
The Perseus cluster of galaxies (Abell 426) is one of the best studied
clusters, due to its proximity ($z=0.018$, $1'$ corresponds to $\sim$
30 kpc for $H_0=50~km~s^{-1}~Mpc^{-1}$) and brightness. Detailed X--ray
images were obtained with the Einstein IPC (Branduardi--Raymont et
al. 1981) and HRI (Fabian et al. 1981) and the ROSAT PSPC (Schwarz et
al. 1992, Ettori, Fabian, White 1999) and HRI (B\"{o}hringer et al. 1993;
see also Heinz et al. 1998). The cluster has a prominent X--ray surface
brightness peak at its center along with cool gas, which is usually
interpreted as due to the pressure induced flow of gas releasing its thermal
energy via radiation. The cooling flow is centered on the active galaxy
NGC1275, containing a strong core-dominated radio source (Per A, 3C 84)
surrounded by a lower surface brightness halo (e.g. Pedlar et
al. 1990, Sijbring 1993). Analysis of the ROSAT HRI observations of
the central arcminute has shown that the X-ray emitting gas is
displaced by the bright radio emitting regions (B\"{o}hringer et
al. 1993), suggesting that the cosmic ray pressure is at least
comparable to that of the hot intracluster gas. Many other studies
explored correlations of X-ray, radio, optical, and ultraviolet
emission (see e.g. McNamara, O'Connell \& Sarazin, 1996 and references
therein). In this contribution, we discuss asymmetric structure in the
X--ray surface brightness within $\sim$ 5 arcminutes of NGC 1275
and suggest that buoyant bubbles of relativistic plasma may be
important in defining the properties of this structure.

\section{IMAGES}
\begin{figure}
\plotone{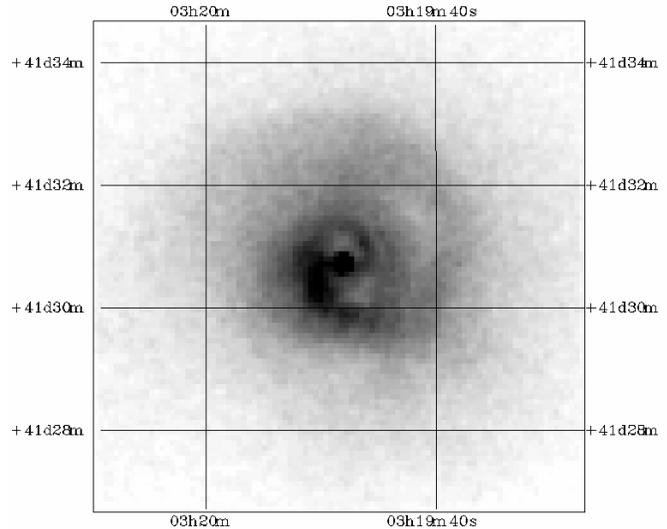}
\caption[]{The $8' \times 8'$ subsection of the ROSAT HRI image  
convolved with a $3''$ Gaussian. The image is centered at NGC 1275. Two
X--ray minima immediately to the north and south of NGC 1275 coincide
(B\"{o}hringer et al. 1993) with bright lobes of radio emission at 332 MHz.
Another region of reduced brightness ($\sim
1.5'$ to the north--west from NGC 1275) was detected earlier in Einstein IPC
and HRI images (Branduardi--Raymont et al. 1981, Fabian et al. 1981).}
\label{raw}
\end{figure}

\begin{figure}
\plotone{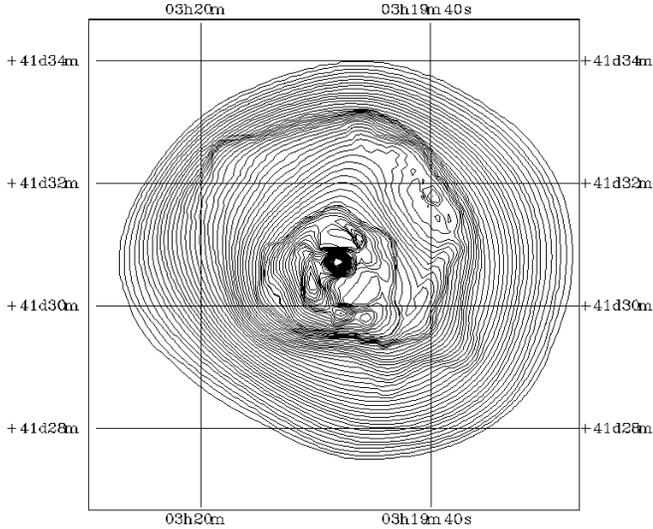}
\caption[]{The same image as in Fig.1 adaptively smoothed using the
wavelet--based procedure of Vikhlinin, Forman, Jones (1996). Contours are
plotted with multiplicative increments of 1.05.}   
\label{wv}
\end{figure}

The longest ROSAT HRI pointing towards NGC~1275 was made in August 1994
with a total exposure time of about 52 ksec. The $8' \times 8'$ subsection of
the HRI image, smoothed with a $3''$ Gaussian, is shown in
Fig.~\ref{raw}. The image is centered  at NGC 1275. Two X--ray minima
immediately to the north and south of NGC 1275 coincide (B\"{o}hringer et
al. 1993) with bright lobes of radio emission at 332 MHz, mapped with the
VLA by Pedlar et al. (1990). Another region of reduced brightness ($\sim
1.5'$ to the north--west from NGC 1275) was detected earlier in Einstein IPC
and HRI images (Branduardi--Raymont et al. 1981, Fabian et al. 1981). It was
suggested that reduced brightness in this region could be due to a
foreground patch of a photoabsorbing material or pressure driven asymmetry
in the thermally unstable cooling flow (Fabian et al. 1981). The
complex shape of the X-ray surface brightness is much more clearly seen in
Fig.~\ref{wv} which shows the same image, adaptively smoothed using the
procedure of Vikhlinin, Forman, Jones (1996). The ``compressed'' isophotes in the
figure delineate a complex spiral--like structure. Comparison of Fig.~\ref{wv}
and Fig.~\ref{raw} shows that the same structure is present in both images,
i.e. it is not an artifact of the adaptive smoothing procedure. 

\begin{figure}
\plotone{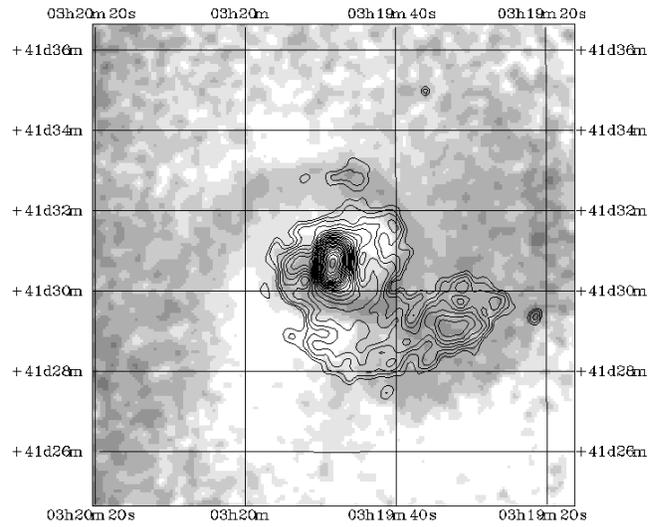}
\caption[]{The $12' \times 12'$ subsection of the HRI image divided by the
azimuthally averaged  surface brightness profile at a given distance from
NGC 1275 and convolved with a $6''$ Gaussian. It characterizes the value of
the surface brightness relative to the azimuthally averaged value. The
darker the color, the larger is the value 
(white color corresponds to regions with surface brightness lower than
azimuthally averaged value; dark grey corresponds to region which are $\sim$
30\% brighter than the azimuthally averaged value). Superposed onto the image
are the contours of the radio flux at 1380 MHz (Pedlar et al. 1990). The
radio data have a resolution of $22 \times 22$ arcsec$^2$ and the central
region is not resolved. The compact feature to the west of NGC 1275, visible
both in X--rays and radio is the radio galaxy NGC 1272. 
}  
\label{radio}
\end{figure}

In order to estimate the amplitude of the substructure relative to
the undisturbed ICM, we divided the original image (Fig.\ref{raw}) by the
azimuthally averaged radial surface brightness profile.  The resulting
image, convolved with 
the $6''$ Gaussian is  
shown in Fig.\ref{radio}. The regions having surface brightness higher than 
the azimuthally averaged value appear grey in this image and form a long
spiral--like structure starting near the cluster center and ending $\sim 5'$ from
the center to the south--east. Of course the appearance of the excess
emission as a ``spiral'' strongly depends on the choice of the ``undisturbed''
ICM model (which in the case of Fig.\ref{radio} is a symmetric distribution
around NGC 1275). Other models would imply different shapes for the regions
having excess emission. In particular a substantial part of the subtructure
seen in Fig.\ref{raw} and \ref{wv} can be accounted for by a model
consisting of a sequence of ellipses with varying centers and position
angles (e.g. using the IRAF procedure {\bf ellipse} due to Jedrzejewski,
1987). Nevertheless the image shown in  Fig.\ref{radio} provides a convenient
characterization of the deviations of the X--ray surface brightness relative
to the 
azimuthally averaged value. Comparison of Fig.\ref{radio} and Fig.\ref{raw},
\ref{wv} allows one to trace all features visible in Fig.\ref{radio} back
to the original image. 

Superposed onto the image shown in Fig.\ref{radio} are the contours of
the radio image of 3C 84 at 1380 MHz (Pedlar et al.  1990). The radio
image was obtained through DRAGN atlas ({\bf
http://www.jb.man.ac.uk/atlas} edited by J. P. Leahy, A . H.  Bridle,
and R. G. Strom). In this image, having a resolution of $22 \times 22$
arcsec$^2$, the central region is not resolved (unlike the higher
resolution image of the central area used in B\"{o}hringer et
al. 1993) and it does not show features, corresponding to the
gas--voids north and south of the nucleus.  The compact feature to the
west of NGC 1275, visible both in X--rays and radio, is the radio
galaxy NGC 1272. Fig.\ref{radio} hints at possible relations between
some prominent features in the radio and X--rays. In particular, the
X--ray underluminous region to the North--West of NGC 1275
(Branduardi--Raymont et al. 1981, Fabian et al. 1981) seems to
coincide with a ``blob'' in radio. A somewhat better correlation is
seen if we compare our image with the radio map of Sijbring (1993),
with its better angular resolution), but again the correlation is not
one to one\footnote{Note, that some random correlation is expected
since both the X--ray and radio emission are asymmetric and centrally
concentrated around NGC 1275}. A similar partial correlation of
X--rays and radio images also was found for another well studied
object -- M87 (B\"{o}hringer et al. 1995). For M87, the relatively
compact radio halo surrounds the source and some morphological
similarities of the X--ray and radio images are observed. Gull and
Northover (1973) suggested that buoyancy plays an important role in
the evolution of the radio lobes. B\"{o}hringer et al. (1993, 1995)
pointed out that buoyant bubbles of cosmic rays may affect the X--ray
surface brightness distribution in NGC 1275 and M87. Below we
speculate on the hypothesis that this mechanism is operating in both
sources and the disturbance of X--ray surface brightness is related,
at least partly, with the activity of an AGN in the past.

\section{EVOLUTION OF THE OLD RADIO LOBES}
The complex substructure of the X--ray emission in the Perseus cluster is
seen at various spatial scales. At large
scales (larger than $\sim 10'$ -- $20'$), excess emission to the east of NGC
1275 was observed in the HEAO--2 IPC and ROSAT images (Branduardi--Raymont
et al. 1981, Fabian et al. 1981, Schwarz et al. 1992, Ettori, Fabian, White
1999). Schwarz et al. (1992), using ROSAT PSPC data, found that the
temperature is lower in this region and suggested  that there is a
subcluster projected on the A426 cluster and merging with the main cluster.
At much smaller scales ($\le 1'$), there are two X--ray minima
(symmetrically located to the north and south of NGC 1275) which
B\"{o}hringer et al. (1993) explained as due to the displacement of the
X--ray emitting gas by the high pressure of the radio emitting plasma
associated with the radio lobes around NGC 1275. As is clear from
Fig.\ref{raw},\ref{wv} at intermediate  scales (arcminutes),  
substructure is also present. We concentrate below on the possibility that
at these spatial scales, the disturbed X--ray surface brightness
distribution is affected by the bubbles of radio emitting plasma, created by
the jets in the past and moving away from the center due to buoyancy. 

Recently Heinz, Reynolds \& Begelman (1998) argued that the
time-averaged power of the jets in NGC 1275 exceeds $\sim
10^{45}~ergs~s^{-1}$. This conclusion is based on the observed
properties (in particular -- sharp boundaries) of the X--ray cavities
in the central $1'$, presumably inflated by the relativistic particles
of the jet. Such a high power input is comparable to the total X--ray
luminosity of the central $6'$ region (i.e. $\sim$ 200 kpc) around NGC
1275. If the same power is sustained for a long time (e.g. cooling
time of the gas $\sim 10^{10}$ years at a radius of 200 kpc) then the
entire cooling flow region could be affected. Following Gull and
Northover (1973) we assume that buoyancy (i.e. Rayleigh--Taylor
instability) limits the growth of the cavities inflated by the
jets. After the velocity of rise due to buoyancy exceeds the expansion
velocity, the bubble detaches from the jet and begins rising. As we
estimate below, for the jet power of $\sim 10^{45}~ergs~s^{-1}$ the
bubble at the time of separation from the jet should have a size
$\sim$10--20 kpc ($\le 1'$). The subsequent evolution of the bubble
may resemble the evolution of a powerful atmospheric explosion or a
large gas bubble rising in a liquid (e.g. Walters and Davison 1963,
Onufriev 1967, Zhidov et al. 1977). If the magnetic field does not
provide effective surface tension to preserve the quasi--spherical
shape of the bubble, then it quickly transforms into a torus and
mixes with the ambient cooling flow gas.  The torus keeps rising until
it reaches the distance from the center where its density (accounting
for adiabatic expansion) is equal to the density of the ambient
gas. Since the entropy of the ICM rises with distance from the center
in the cooling flow region, the torus is unlikely to travel a very
large distance from the center. Then the torus extends in the lateral
direction in order to occupy the layer having a similar mass density.
Below we give order of magnitude estimates characterizing the
formation and evolution of the bubble.

For simplicity we assume a uniform ICM in the cluster center, characterised by
the density $\rho_0$ and pressure $P_0$. The bubble is assumed to be
spherical. During the initial phase (Scheuer 1974,
Heinz et al., 1998) jets with a power $L$ inflate the cocoon 
with
relativistic plasma and surrounded by a shell of the compressed ICM.
The expansion is supersonic and from dimensional arguments
it follows that the radius of the bubble $r$ as a function of time $t$ is
given by the expression 
\begin{eqnarray}        \label{r1}
r=C_1 \left (\frac{L}{\rho_0}t^3 \right )^{1/5} 
\end{eqnarray} 
where $C_1$ is a numerical constant (see e.g. Heinz et al., 1998 for a more
detailed treatment). At a later stage, expansion slows and becomes
subsonic. The evolution of the bubble radius is then given by the expression
\begin{eqnarray}        \label{r2}
r=C_2 \left (\frac{L}{P_0}t \right )^{1/3}=\left (\frac{3}{4\pi}\frac{\gamma-1}{\gamma} \right )^{1/3}\left (\frac{L}{P_0}t \right )^{1/3}
\end{eqnarray} 
where $\gamma$ is the adiabatic index of the relativistic gas in the bubble
(i.e. $\gamma=4/3$). The above equation follows from
the energy conservation law, if we equate the power of the jet with
the  change of internal energy plus the work done by the expanding gas at constant pressure:
$\frac{\gamma}{\gamma-1}P_04\pi r^2 \dot{r}=L$. The expansion velocity is
then simply the time derivative of equation (\ref{r1}) or (\ref{r2}). 

The velocity at which the bubble rises due to buoyancy can be estimated as  
\begin{eqnarray}        \label{v1}
v_b=C_3 \sqrt{\frac{\rho_{0}-\rho_{r}}{\rho_{0}+\rho_{r}}rg}=C_3\sqrt{\frac{r}{R}}\sqrt{\frac{GM}{R}}=C_3\sqrt{\frac{r}{R}}v_K
\end{eqnarray} 
where $C_3$ is a numerical constant of order unity, $\rho_{r}$ is
the mass density of the relativistic gas in the bubble, $g$ is the
gravitational acceleration, $R$ is the distance of the bubble from the
cluster center, $M$ is the gravitating mass within this radius and,
$v_K$ is the Keplerian velocity at this radius. In equation (\ref{v1})
we assumed that $\rho_{r}\ll\rho_{0} $ and therefore replaced the
factor $\frac{\rho_{0}-\rho_{r}}{\rho_{0}+\rho_{r}}$ (Atwood number)
with unity.  The presently observed configuration of the bubbles on
either side of NGC 1275 suggests that $r\sim R$. Assuming that such a
similar relation is approximately satisfied during the subsequent
expansion phase of the bubble we can further drop the factor
$\sqrt{\frac{r}{R}}$ in equation (\ref{v1}). Thus as a crude estimate
we can assume that $v_b\sim C_3~v_K$ ($C_3\sim 0.5$ is a commonly
accepted value for incompressible fluids).  Following Ettori, Fabian,
and White (1999) we estimate the Keplerian velocity taking the gravitating
mass profile as a sum of the Navarro, Frenk and White (1995) profile
for the cluster and a de Vaucouleurs (1948) profile for the
galaxy. For the range of parameters considered in Ettori, Fabian,
White (1999), the Keplerian velocity between a few kpc and $\sim$ 100 kpc
falls in the range 600-900 km/s. We can now equate the expansion
velocity (using equation (\ref{r2}) for subsonic expansion) and the
velocity due to buoyancy in order to estimate the parameters of the
bubble when it starts rising:
\begin{eqnarray}        \label{tb}
\nonumber t_b= \left (\frac{1}{36\pi}\frac{\gamma-1}{\gamma}\right
)^\frac{1}{2}  \left (\frac{L}{P_0}\right
)^\frac{1}{2}\left (\frac{1}{C_3 v_K}\right )^\frac{3}{2} \approx 
\\ 1.6~10^7~ \left (\frac{L}{10^{45}}\right )^\frac{1}{2} 
\left (\frac{P_0}{2~10^{-10}}\right )^{-\frac{1}{2}} 
\left (\frac{v_K}{700}\right )^{-\frac{3}{2}} ~ years
\end{eqnarray}  
\begin{eqnarray}        \label{rb}
\nonumber r_b=\left ( \frac{L}{P_0C_3v_k} \frac{\gamma -1}{\gamma}
\frac{1}{4\pi}\right )^\frac{1}{2}\approx \\ 
17~ 
\left (\frac{L}{10^{45}}\right )^\frac{1}{2} 
\left (\frac{P_0}{2~10^{-10}}\right )^{-\frac{1}{2}} 
\left (\frac{v_K}{700}\right )^{-\frac{1}{2}} ~ kpc
\end{eqnarray}  
Here $t_b$ and $r_b$ are the duration of the expansion phase and the radius of
the bubble respectively. 
In the above equation we neglected the contribution to the radius (and time)
of the initial supersonic expansion phase. Thus for $L\sim 10^{45}~erg/s$
and for $P_0=2~10^{-10}~erg~cm^{-3}$ (B\"{o}hringer et al. 1993) we expect
$r_b\sim17~kpc$, which approximately corresponds to the size of the X--ray
cavities reported by B\"{o}hringer et al. (1993). If, as suggested by Heinz et
al. (1998), the jet power is larger than $10^{46}~erg~s^{-1}$ then the bubble
size will exceed 50 kpc ($> 1'$) before the buoyancy velocity exceeds 
the expansion velocity.  Of course these estimates of the expanding
bubble are based on many simplifying assumptions (e.g. constant
pressure assumption in equation (2)). In a subsequent publication we
consider the expansion of the bubble in  more realistic density and
temperature profiles expected in cluster cooling flows.

According to e.g. Walters and Davison (1963), Onufriev (1967), Zhidov
et al. (1977), a large bubble of light gas rising through much heavier
gas under a buoyancy force will quickly transform into a rotating
torus, which consists of a mixture of smaller bubbles of heavier and
lighter gases. This transformation occurs on times scales of the
Rayleigh--Taylor instability (i.e. $t\sim r_b/v_b\sim t_b$) and during
this transformation the whole bubble changes its distance from the
center by an amount $\sim r_b$. The torus then rises until its
average mass density is equal to the mass density of the ambient
gas. The rise is accompanied by adiabatic expansion and further mixing
with the ambient gas.  Accounting for adiabatic expansion the mass
density of the torus $\rho_t(R)$ will change during the rise according
to
\begin{eqnarray}        \label{dens}
\rho_t(R)=\rho_0 \frac{\phi}{(1-\phi) \left ( \frac{P_0}{P(R)} \right )^{1/\gamma_{cr}} +
\phi \left ( \frac{P_0}{P(R)} \right ) ^{1/\gamma_{th}}} 
\end{eqnarray}
where $P(R)$ is the ICM pressure at a given distance from the center, $\phi$
is volume fraction of the ambient ICM gas mixed with the relativistic plasma
at the stage of torus formation, $\gamma_{cr}$ and $\gamma_{th}$ are the
adiabatic indices of the relativistic plasma and the ICM. Note that in
equation (\ref{dens}) we (i) neglected further mixing with the ICM during
the rise of the torus and (ii) mixing was assumed to be macroscopic (i.e.
separate bubbles of the relativistic plasma and ICM occupy the volume of the
torus). The equilibrium position of the torus can be found if we
equate the torus
density $\rho_t(R)$ and the ICM density $\rho(R)$ and solve this equation for
$R$. We consider two possibilities here. One possibility is to assume that in
the inhomogeneous cooling flow, the hot phase is almost isothermal and gives
the dominant contribution to the density of the gas. Adopting the
temperature of 
$kT=6$ keV for the hot phase and using the same gravitational potential as
above, one can conclude that if a roughly equal amount (by volume) of the
relativistic plasma and the ambient gas are mixed (i.e. $\phi\sim0.5$),
during the formation of the torus, then it could rise 100-200 kpc
 before reaching an equilibrium position. Accounting for additional
mixing will lower this estimate. Alternatively we can adopt the model of a
uniform ICM with the temperature declining towards the center (e.g.
temperature is decreasing from 6 keV at 200 kpc to 2 keV at 10 kpc). Then
for the same value of mixing ($\phi\sim0.5$) the equilibrium position will
be at the distance of $\sim$60 kpc from NGC 1275. Once at this distance the
torus as a whole will be in equilibrium and it will further expand laterally
in order to occupy the equipotential surface at which the density
of the ambient gas is equal to the torus density. If the cosmic rays and
thermal gas within the torus are uniformly mixed (or a magnetic field binds
the blobs of thermal plasma and cosmic rays), the torus will not move 
radially. If, on the contrary, separate (and unbound) blobs of 
relativistic plasma exist then they will still be buoyant, but since their
size is now much smaller than the distance from the cluster center the
velocity of their rise will be much smaller than the Keplerian velocity.
Analogously overdense blobs (with uplifted gas) may then (slowly) fall back
to the center.  

We now consider how radio and X--ray emission from the torus evolve with
time. Duration of the rise phase of the torus will be at least several times
longer than the time of the bubble formation (see eq. (\ref{tb})), since the
velocity of rise is a fraction of the Keplerian velocity (see e.g. Zhidov et
al., 1977), i.e.,  $t_{rise} \ge 10^8$ years. Adiabatic expansion and change of
the transverse size of the torus in the spherical potential tend to further
increase this estimate. Even if we neglect energy losses
of the  relativistic electrons due to adiabatic expansion we can
estimate an upper
limit on the electron lifetime due to synchrotron and inverse Compton (IC) 
losses.  
\begin{eqnarray}        \label{lifet}
t=5~10^8~ \left ( \frac{\lambda}{20~cm} \right )^{1/2} \left ( \frac{B}{\mu
G} \right )^{1/2}
\left ( \frac{B_t}{\mu G} \right )^{-2}~~years
\end{eqnarray}
where $\lambda$ is the wavelength of the observed radio emission, $B$
is the strength of the magnetic field, $\frac{B_t^2}{8\pi}$ is the
value characterizing the total energy density of the magnetic field
and cosmic microwave background. This life time (of the electrons
emitting at a given frequency) will be longest if the energy density
of the magnetic field approximately matches the energy density of the
microwave background, i.e., $B\sim3.5\mu G$. Then the maximum life
time of the electrons producing synchrotron radiation at 20 cm is
$\sim 5~10^7$ years.  This time is comparable to the time needed for
the torus to reach its final position. Therefore, the torus could be
either radio bright or radio dim during its evolution. If no
reacceleration takes place, then the torus will end up as a radio dim
region. We note here that, although the electrons may lose their energy
via synchrotron and IC emission, the magnetic field and especially
relativistic ions have a much longer lifetime (e.g. Soker \& Sarazin
1990, Tribble 1993) and will provide pressure support at all stages of
the torus evolution.

As we assumed above, the bubble detaches from the jet when the expansion
velocity of the bubble is already subsonic. This means that there will be no
strongly compressed shell surrounding the bubble and the emission measure along the
line of sight going through the center of the bubble will be smaller than
that for the undisturbed ICM, i.e., at the moment of detachment the bubble
appears as an X--ray dim region. The X--ray brightness of the torus
during final stages of evolution (when the torus has the same mass density
as the ambient ICM) depends on how the relativistic plasma is mixed with the
ambient gas (B\"{o}hringer et al. 1995). If mixing 
is microscopic (i.e. relativistic and thermal particles are uniformly
mixed over the torus volume on  spatial scales comparable with
the mean free
path) then the emission measure of the torus is the same as that for a similar
region of the undisturbed ICM. Since part of the pressure support in the
torus is provided by magnetic field and cosmic rays then the temperature of the
torus gas must be lower than the temperature of the ambient gas
(B\"{o}hringer et al. 1995). Thus emission from the torus will be softer
than the emission from the ambient gas.

If, on the contrary, mixing is macroscopic (i.e. separate bubbles 
of relativistic and thermal plasma occupy the volume of the torus), then the
torus will appear as an X--ray bright region (the average density is the
same as of ICM, but only a fraction of the torus volume is occupied by the thermal
plasma). For example, if half of the torus volume is occupied by the bubbles
of the relativistic plasma then the emissivity of the torus will be a factor
of 2 larger than that of the ambient gas. The X--ray emission 
of the torus is again expected to be softer than the emission of the ambient gas
for two reasons (i) gas uplifted from the central region has lower entropy
than the ambient gas and therefore will have lower temperature when maintaining
pressure equilibrium with the ambient gas (ii) gas, uplifted from the central
region, can be multiphase with stronger density contrasts between
phases than the ambient gas and as a result a dense, cooler phase would
give a strong contribution to the soft emission. Cosmic rays may heat the gas,
but at least for the relativistic ions, the time scale for energy transfer is
very long (comparable to the Hubble time). Trailing the torus could be the
filaments of cooling flow gas dragged by the rising torus in a similar
fashion as the rising (and rotating) torus after an atmospheric
explosion drags the air in the form of a skirt.  

We note here that the morphology predicted by such a picture is very similar
to the morphology of the ``ear--like'' feature in the radio map of M87,
reported by B\"{o}hringer et al. (1995). The ``ear'' could be a torus viewed
from the side. The excess X--ray emission trailing the radio feature
could then be due to the cooling flow gas uplifted by the torus from the
central region. For Perseus the X--ray underluminous 
region to the North--West of NGC 1275 could have the same origin (i.e.
rising torus). In fact, the whole ``spiral'' structure seen in
Fig.\ref{wv} could 
be the remains of one very large bubble (e.g. with the initial size of the
order of arminutes -- corresponding to a total jet power of $\sim
10^{46}~erg~s^{-1}$) inflated by the nucleus over a period of $10^8$
years. Alternatively 
multiple smaller bubbles, produced  at different periods may contribute to
the formation of the X--ray feature. If the jets maintain their direction over
a long time then a quasi--continuous flow of bubbles will tend to mix the
ICM in these directions uplifting the gas from the central region to larger
distances. If the jet direction varies (e.g. precession of the jet on a
timescale of $10^8$ years) then a complex pattern of disturbed X--ray and
radio features may develop.

\section{ALTERNATIVE SCENARIOS}
Of course there are other possible explanations for disturbed X--ray surface
brightness. We briefly discuss a few alternative scenarios below.

Assuming that the undisturbed ICM is symmetric around NGC 1275 (as was assumed
in Fig.\ref{radio}) one may try to attribute the observed spiral-shaped
emission to the gas stripped from an infalling galaxy or group of
galaxies. Stripped gas (if denser and cooler than the ICM) will be
decelerated by ram pressure and will fall toward the center of the
potential, producing spiral--like structure. Rather narrow and long
features tentatively associated with stripped gas were observed e.g.,
for the NGC 4921 group in Coma (Vikhlinin et al. 1996) and NGC
4696B in the Centaurus cluster (Churazov et al. 1999). We note here
that to prevent stripping at much larger radii, the gas 
must be very dense (e.g., comparable to the molecular content of a
spiral galaxy). A crude estimate of the gas mass needed to produce the
observed excess emission (assuming a uniform cylindrical feature with
a length of 200 kpc and radius of 15 kpc, located 60 kpc away from 
NGC1275) gives values of the order of a few$\times~10^{10}$ --
$10^{11}~M_\odot$. Here we adopted a density for the undisturbed ICM of
$\sim 10^{-2}~cm^{-3}$ at this distance from NGC 1275 following the
deprojection analysis of Fabian et al. (1981) and Ettori, Fabian, and White
(1999). The factor of two higher density within the feature will
cause a $\sim$ 20--40\%
excess in the surface brightness. In the above estimate for the mass of hot
gas in the filament, it is assumed that this medium is approximately
homogeneous and in ionization equilibrium. If the medium is very clumpy, the
radiative emission of the plasma would be enhanced and this would result in an
overestimate of the relevant gas mass. Such clumps should be easily
seen with the high angular resolution of Chandra. Also if the medium consists of
turbulently mixed hot and cold plasma, the very efficient excitation
of lines in cold ions by hot electrons could lead to enhanced
radiation (see e.g. B\"ohringer and Fabian 1989, Table 4) which may
lead to an overestimate  of the gas mass by up to an order of magnitude.
The signature of this effect is a strongly line dominated spectrum,
(see e.g. B\"ohringer and Hartquist, 1987) which could be tested by
Chandra or XMM, in particular for the important iron L-shell lines.
Thus it is possible that the inferred gas mass could be lower by up to
an order of magnitude which makes the stripping scenario more likely
and future observations with the new X-ray observatories
can help to differentiate between these interpretations. 

As was suggested by Fabian et al. (1981) a large scale pressure--driven
asymmetry may be expected in a thermally unstable cooling flow. This is
perhaps the most natural explanation which does not invoke any additional
physics. The same authors gave an estimate of the amount of neutral
gas needed to explain the NW dip due to photoabsorption: excess
hydrogen column density around $10^{22}~cm^{-2}$ is required to
suppress the soft count rate in this region.  

Yet another possibility is that the motion of NGC 1275 with respect to
the ICM causes the observed substructure. As pointed out in
B\"{o}hringer et al. (1993), NGC 1275 is perhaps oscillating at the
bottom of the cluster potential well causing the excess emission $1'$ to the
east of the nucleus. Since the X-ray surface brightness peak is well
centered on NGC 1275, it is clear that the galaxy drags the central part of the
cooling flow as it moves in the cluster core. At a distance larger
than 2-3 arcminutes from NGC 1275, the cluster potential dominates
over the potential of the galaxy. The gas at this distance should be very
sensitive to the ram pressure of the ambient cluster gas and might give rise
to the asymmetric (and time dependent) features. 

The motion of NGC~1275 could also contribute to the X-ray structure
through the formation of a ``cooling wake'' (David et al. 1994). If
NGC~1275 is moving significantly, then inhomogeneities in the cooling
gas would be gravitationally focussed and compressed into a wake. The
wake would mark the, possibly complex, motion of NGC~1275, as it is
perturbed by galaxies passing through the cluster core. Such a feature
would be cool, since it arises from overdense concentrations of gas.

Finally, one can assume that cooling gas may have some angular
momentum (e.g., produced by mergers) and the observed spiral structure
simply reflects slow rotation of the gas combined with non-uniform
cooling. Following Sarazin et al. (1995), one can assume that this gas
will preserve the direction of its angular momentum and
that this infalling material would eventually feed an AGN --
NGC~1275. One then might expect the radio jets to be aligned
perpendicular to the rotation plane of the gas. At first glance, the
``spiral'' feature appears approximately face-on, suggesting that jets
should be directed along the line of sight as indeed is derived from
the radio observations (see Pedlar et al. 1990).

\section{CONCLUSIONS}
The X--ray surface brightness around NGC 1275 (dominant galaxy of the
Perseus cluster) is perturbed at various spatial scales. We suggest that
on arcminute scales, the disturbance is caused by bubbles
of relativistic plasma, inflated by jets during the past $\sim 10^8$ years.   
Overall evolution of the buoyant bubble will resemble the evolution of a hot
bubble during a powerful atmospheric explosion. Colder gas from the central
region of the cooling flow may be uplifted by the rising bubbles and (in the
case of continuous jet activity) may make several cycles (from the center to
the outer regions and back) on time scales comparable to the cooling
time of the gas in the cooling flow.

A very important result that can be inferred from this model
is the total power output of the nuclear energy source in 
NGC 1275 in the form of relativistic plasma. This energy
release averaged over a time scale of about $3~10^7$ to $10^8$
years is estimated as a function of the inflation time of the
central radio lobes, the rise time of the inflated bubbles
due to buoyancy forces, and the actual size of the central 
bubbles. A geometrically simplified model yields a power 
output on the order of $10^{45}$ erg s$^{-1}$. This is comparable with
the energy lost at the same time by thermal X-ray radiation from the
entire central cooling  flow region. This raises the question, where
does all this energy go, especially if the energy release is persistent over a
longer epoch during which the relativistic electrons can lose their
energy by radiation, but  the energy in protons and in the
magnetic field is mostly conserved. The complicated X-ray
morphology discussed in this paper may indicate long
lasting nuclear activity, if we interpret the peculiar structure
in the X-ray surface brightness as remnants of decaying  radio
lobe bubbles.

Detailed measurements of the morphology of the X--ray structure and
the temperature and abundance distribution with Chandra and XMM may
test this hypothesis. The gas uplifted from the central region is
expected to be cooler than the ambient gas and to have an abundance of
heavy elements typical of the innermost region. If cosmic rays are
mixed with the thermal gas, then the pressure, as derived from X--ray
observations, may be lower than the pressure of the ambient gas.

\acknowledgements

We thank the referees for several helpful comments and
suggestions. We are grateful to Nail Inogamov and Nail Sibgatullin
for useful discussions.
This research has made use of data obtained through the High Energy
Astrophysics Science Archive Research Center Online Service, provided
by the NASA/Goddard Space Flight Center.  W. Forman and C. Jones
acknowledge support from NASA contract NAS8-39073.

\end{document}